\documentclass[12pt]{iopart}
\usepackage{graphicx}
\usepackage{epstopdf} 
\begin{document}

\title[The LEGO Representation and Origin of Mass Functions]{The Pedagogical Representation of Mass Functions with LEGO and their Origin}

\author{Stefan J. Kautsch$^1$, Dimitri Veras$^{2,3}$, Kyle K. Hansotia$^1$}
\address{$^1$ Department of Chemistry and Physics, Halmos College of Arts and Sciences, Nova Southeastern University, Fort Lauderdale, FL 33314, USA}

\address{$^2$ Centre for Exoplanets and Habitability, University of Warwick, Coventry CV4 7AL, UK}
\address{$^3$ Department of Physics, University of Warwick, Coventry CV4 7AL, UK}

\ead{skautsch@nova.edu}
\vspace{10pt}
\begin{indented}
\item[]This is the version of the article before submission and peer review. The published version is Kautsch, Veras, \& Hansotia 2021, {\it European Journal of Physics}, {\bf 42}, 035605 (\verb"https://doi.org/10.1088/1361-6404/abe75c"). The published version has the same content and results, and includes more educational references. IOP Publishing Ltd is not responsible for any errors or omissions in this version of the manuscript or any version derived from it.

\end{indented}

\begin{abstract}
We promote the teaching of mass functions as an integral part of an interdisciplinary science education. Mass functions characterize the frequency distributions of objects with different masses on all cosmic scales. We intend to enhance experiential learning of this concept with a creative LEGO brick experiment for a diverse student audience. To our surprise, the LEGO mass function is not only qualitatively but also quantitatively comparable to mass functions found across the Universe. We also discuss the relation between gravitation and mass distributions as a possible explanation for the continuity of the universal mass function.
\end{abstract}

%
\vspace{2pc}

\section{Introduction}

The teaching of the concept of mass functions aims to broaden the understanding of mass distributions in the cosmos. The target audiences are students of introductory physics and astronomy courses in higher education. We also want to engage students in general, undergraduate STEAM (science, technology, engineering, arts, math) courses to embrace this topic, because similar distribution concepts can be found in other natural sciences, mathematics, social sciences, arts, applied sciences, etc. In this article, we describe an inexpensive experiment which can improve the comprehension of this concept and can be performed in a variety of creative ways. We also explain the potential origin of mass functions in the cosmos.

Mass is one of the most important physical properties of objects in space. For instance, it determines the lifetime of stars, the dark matter content of galaxies, and determines how planets can harbor life as we know it, or simply how heavy objects are. Therefore, number and frequency distributions of masses are of great interest for astrophysical research. If the mass distributions are known, then they help to explain structure formation and the subsequent evolution of matter in the Universe. 

Astronomical objects (planetary objects, stars, galaxies, dark matter halos, etc.) can have a wide variety of individual masses. A fundamental observation is that objects with low masses are far more abundant than massive ones within a unit of space and of a given kind. For instance, our Milky Way galaxy contains many more low-mass stars compared to only very few massive stars.  Moreover, the same principle can be seen when comparing objects of different kinds (e.g., more asteroids than planets, more planets than stars, more stars than galaxies).

This number frequency distribution of masses of many objects can be mathematically quantified with a mass function. Thus, the mass function is a fundamental law of mass distribution in the Universe. Despite its importance, this concept is rarely presented in introductory courses and is barely discussed in popular physics and astronomy textbooks, and is even lesser known in other sciences. A few advanced, general astronomy textbooks (e.g., \cite{kut03}, \cite{co17}) do actually cover mass functions but these are limited to the context of stellar astrophysics. 

We believe that the concept of mass function is important because of its universality. This topic should be an integral part of the introductory physics, astronomy, as well as the STEAM curriculum.  Moreover, we also think that this concept has philosophical implications for how everything is connected in the cosmos and can inspire students and their educators to advance their understanding of the Universe. We present an experiment using LEGO bricks in order to promote easy visualization and application for undergraduate university and college lectures and classes, laboratory courses, and independent studies. The mass function is introduced in chapter \ref{ch1} below. In chapters \ref{ch2} and \ref{ch3} we present the LEGO visualization experiment; and further discuss the concept in chapter \ref{ch4}.

\subsection{Mass Function}\label{ch1}

A mass function is mathematically a power law, also known as scaling law. It is the mathematical model $\Phi(m)$ that fits the number of objects $n$ in different ranges of mass between $m$ and $m+\rmd m$. We use the formalism for mass functions by Hopkins \cite{hop18}. Some publications use other Greek letters like $\xi$ or $\chi$ instead of $\Phi$. The mass function in linear units is:

\begin{equation}
\label{mf}
\frac{\rmd n}{\rmd m}=\Phi(m)=k m^{\alpha}
\end{equation}

$k$ is a constant, i.e., the coefficient of that function. $\alpha$ is the power index, i.e., the slope of the function. The slope determines how quickly the number of high-mass objects is decreasing and low-mass objects is increasing while the proportion of high-to-low mass objects remains constant.

Integration of (\ref{mf}) leads to the useful formula (\ref{nf}) to compute the number of objects with masses between ${m_{1}}$ and ${m_{2}}$ if $k$ and $\alpha$ are known:

\begin{equation}
\label{nf}
n= \int_{m_{1}}^{m_{2}} \Phi(m) \rmd m = k \int_{m_{1}}^{m_{2}} m^{\alpha} \rmd m= \frac{k}{\alpha + 1} (m_{2}^{\alpha + 1} - m_{1}^{\alpha + 1})
\end{equation}

Traditionally, mass functions $\Phi(m)$ (\ref{mf}) have been applied to the distribution of stars for decades. Salpeter \cite{sal55} \cite{kro19} was one of the first who realized the importance of stellar mass functions for understanding stellar evolution. Since then, the goal of such studies is to find the initial mass function (IMF), i.e., the proportion of stars of various masses at the moment of their birth within a unit of space volume and compare it with the present day mass function (PDMF, i.e., the mass distribution of stars at present). The PDMF usually lacks massive stars compared to the IMF, because they have short life spans. Modern IMF studies (see the reviews \cite{cor05} \cite{kro13} \cite{kru14} \cite{hop18} and references therein) are very important for understanding star formation history at different look-back times and environments by comparing the IMF to the PDMF (e.g., \cite{sca86} \cite{cha03} \cite{bov17} \cite{sol19}). 

Individual stars are not the only possible target. Mass functions can be applied to unresolved stellar populations \cite{ken83} \cite{sam07}. Moreover, mass functions are being studied for almost all other types of astronomical objects, such as galaxies and galaxy halos including the enigmatic dark matter \cite{pre74} \cite{sch76} \cite{mof16} \cite{pen19}. Recently, planet hunters have focused on mass functions of planetary bodies \cite{mal15} \cite{bru18} \cite{wag19}.

Zwicky \cite{zwi42} \cite{zwi57} hypothesized that most of the objects in the cosmos follow a single distribution and are therefore universally linked (through a common origin of all mass in the Universe). This idea motivated Binggeli \& Hascher \cite{bin07} to investigate if a single mass function for all astronomical objects exists: the universal mass function. They studied nearly all hierarchies in space, which span 36 orders of magnitude in mass, including asteroids, planets, stars and their remnants, open and globular star clusters, molecular gas and dust clouds, galaxies, galaxy groups and clusters, and  simulated cold dark matter halos. Their surprising result is that all these objects follow a mass function with a constant slope of about $\alpha = -2$ with the simple form $\Phi(m) \propto m^{-2}$ (with an exception for planetary objects with a lower absolute value of the power index and therefore flatter distribution). Because this value is less than $-1$, that confirms that objects with low masses are much more common than massive objects on all cosmic scales. Moreover, the proportion of low-to-high mass objects is nearly constant across all objects in space.

\section{Methods}\label{ch2}

LEGO toy sets naturally exhibit the basic properties to construct a mass function because LEGO pieces, the bricks, come in a variety of masses. Most LEGO sets contain many pieces of low masses and only a few massive bricks, making this toy ideal to show mass frequency distributions \cite{kau20} \cite{han20}. We decided to use the set {\emph{LEGO 70656 NINJAGO garmadon, Garmadon, GARMADON!} \verb"https://www.lego.com/en-us/product/garmadon-garmadon-garmadon-70656" for our experiment, because it contains a shark model like the mascot of our university.

In our experiment, the LEGO bricks of the set were ordered by shape (i.e., mold) and color. The number of bricks per mold and color are classified in the instruction manual of the set. Therefore, we only measured the mass of one brick per mold and color and assumed that the mass difference between identical bricks are negligible for our experiment. Each mass measurement was done with a lab-grade scale and repeated three times. The values are then averaged in order to minimize measurement errors. We defined equidistant mass interval bins and counted the number of bricks within each interval. The number of intervals was naturally limited due to low numbers of massive bricks. Thus, we created tables with four, five, six, and seven bins of mass. 

We use the data and function-plotting program gnuplot \cite{will2}, which uses a nonlinear least-squares Marquardt-Levenberg algorithm for data fitting. The data were fit with the mass function power law (\ref{mf}). The fitting algorithm calculated the final set of values for the constant $k$ and the slope $\alpha$ after a few automatic iterations. No initial values for the parameters were provided for the fitting process.

\section{Results}\label{ch3}
The best fit was produced using the data in six mass intervals. The other mass bin configurations exhibited similar results, but the fitting of six mass bins had the lowest standard errors of the parameters and the best visualization of the mass function. Figure \ref{fig1} shows the mass function as a line in a histogram with the mass of the bricks in grams on the $x$-axis and the numbers of bricks on the $y$-axis. The fitting result values are $\alpha = -2.13 \pm 0.16$; and $k = 266.21 \pm 1.21$. This is also shown in the legend of figure \ref{fig1}.


\begin{figure}[h!]
\begin{center}
\includegraphics[scale=1.7,angle=0]{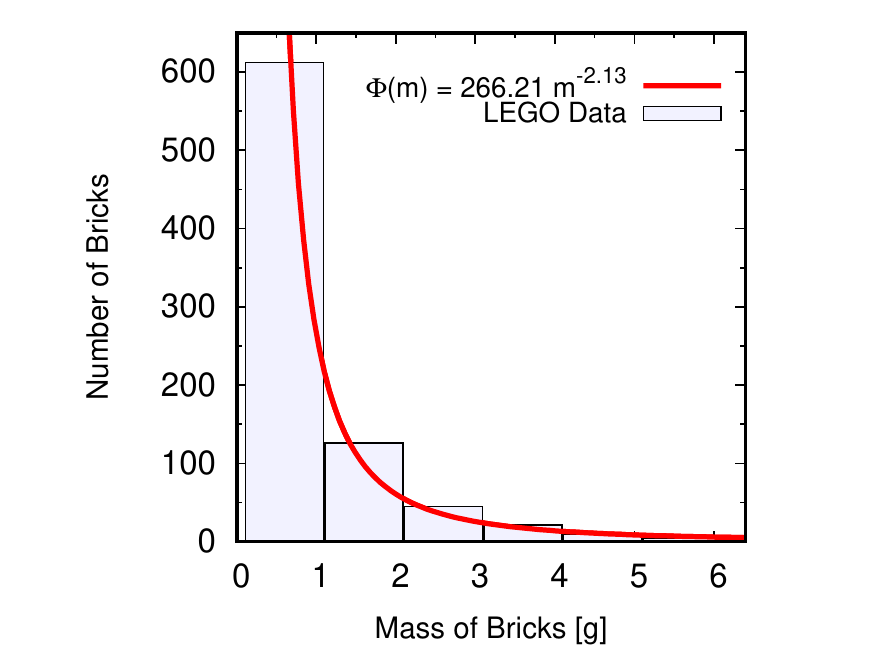}
\caption{The histogram shows the result of our measurements and fitting procedure. The LEGO bricks are sorted into six bins of increasing mass, and fitted with the mass function power law as shown in the legend. \label{fig1}}
\end{center}
\end{figure}

Astronomers have the habit of illustrating mass functions in logarithmic plots. These plots have the advantage of including large ranges of numbers and masses. So, we converted the mass function (\ref{mf}) by taking advantage of the properties of logarithms. We can also test the quality of our fit and confirm the fitting parameters. We use base-ten logarithms, $\log_{10}$, but it can be done for any logarithmic base. The mass function in its logarithmic form is
\begin{eqnarray}\label{logmf}
\log_{10} \Phi(m)&= \log_{10} (k~m^{\alpha})\label{eq2}\\
& = \log_{10} k + \log_{10} m^{\alpha}\label{eq3}\\
& = \alpha \log_{10} m + \log_{10} k \label{eq4}
\end{eqnarray}

(\ref{eq4}) is the equation of a straight line in a $\log_{10}-\log_{10}$ graph. $\alpha$ is still the slope of the line. $\log_{10} k$ is the $y$ intercept of the line at location $\log_{10} m=0$. This conversion enables us to fit the mass function with a linear regression line. Figure \ref{fig2} shows the result of $\log_{10}$(number of bricks) versus $\log_{10}$(mass of bricks in grams). The best-fit parameters are $\alpha = -2.13 \pm 0.16$ and $\log_{10} k = 2.43 \pm 0.08$ and are shown in the legend of figure \ref{fig2}. 

\begin{figure}[h!]
\begin{center}
\includegraphics[scale=1.7,angle=0]{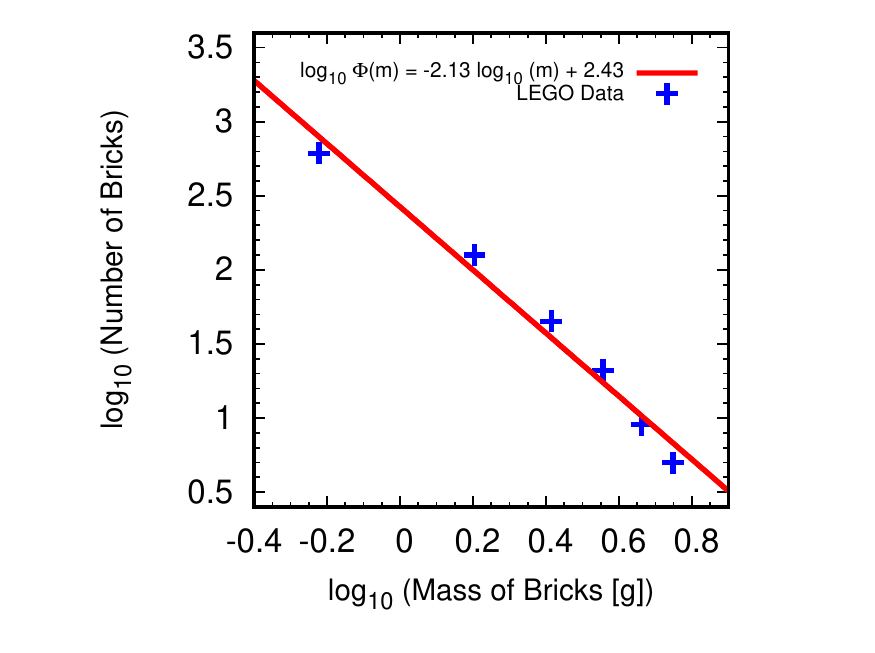}
\caption{Same as figure \ref{fig1} but converted into a linear relationship. This $\log_{10}-\log_{10}$ plot has the advantage to visualize the regression with a straight line. The linear regression line and the best-fit values are shown in the legend. \label{fig2}}
\end{center}
\end{figure}

The fitting algorithm also provided us with the variance of residuals, i.e., the reduced chisquare. It should be around unity if the number of data points is large enough and the fit is excellent.  If this value is lower, then the model function may be too general. Although it is 0.02 in our case, we consider the fit to be good after visual inspection of figure \ref{fig2}. It might be possible to get a value closer to one if our mass distribution is fit with different slopes for different mass intervals. Such broken power laws are commonly used in studies of stellar IMFs. The absolute slope value tends to be larger at the high-mass end of stars. The distribution is flatter---aka bottom-light---for low-mass stars, i.e., the slope has a lower absolute value \cite{mil79} \cite{kro01} \cite{cha03}. We also see slightly different slopes for the low-mass and high-mass intervals in our LEGO fit in figure \ref{fig2}.

\section{Discussion}\label{ch4}

This project can be designed as a classroom demonstration, semester project, an independent study, or a lab experiment with a wide range of creative approaches. We suggest the following main steps: 
1) select a random play set of bricks, discuss selection biases; 2) measure the mass of the individual bricks, discuss measurement uncertainties; 3) select mass intervals, discuss difficulties of interval selections; 4) distribute the bricks into the appropriate mass intervals; 5) count the number of bricks in each interval; 6) plot the data in a diagram; 7) fit the data with function (\ref{mf}); 8) compare the results to the universal mass function and its slope; 9) find similar distributions and discuss potential origins.

\subsection{The Importance of the Slope $\alpha$}

The most important part in the mass function is the slope. It determines the shape of the function. It quantitatively describes the fraction of high-to-low mass objects and how fast the number of low-mass objects increases. Therefore, many modern studies focus to find the value of this slope. Many IMF studies  find $\alpha = -2$ on average for the majority of stars in many galaxies \cite{sal55} \cite{kro01} \cite{cha03} \cite{bas10} \cite{wei13}; although it is highly debated why and whether or not the IMF slope is universally similar \cite{kro13} \cite{hop18}.

Binggeli \& Hascher \cite{bin07} also found a slope of $\alpha = -2$ for their combined mass function, which covers almost all objects in space. They conclude that the function is continuous and the slope is the nearly same for the individual distribution functions as well for the grand total. Everything from stars to galaxy clusters can be fitted with the same universal function and slope $\alpha = -2$. 

It is very surprising that the slope of our LEGO mass function has a similar value, $\alpha = -2.13$. This might be a fortunate coincidence of this particular LEGO set. It needs further investigation of more bricks in other sets to claim that LEGO fits directly into the universal mass function. It shows, however, that LEGO represents the universal mass function not only qualitatively, but also quantitatively. Figure \ref{fig3} shows the visualization of a slope value of $-2$ for LEGO bricks (left) and stars (right). It means that for each star with one solar mass, four stars with $\frac{1}{2}$ the mass of the Sun exist, and 16 stars with $\frac{1}{4}$ of the mass of the Sun exist and so on. The LEGO distribution is the same but for bricks instead of stars. 

\begin{figure}[h!]
\begin{center}
\includegraphics[scale=0.5,angle=0]{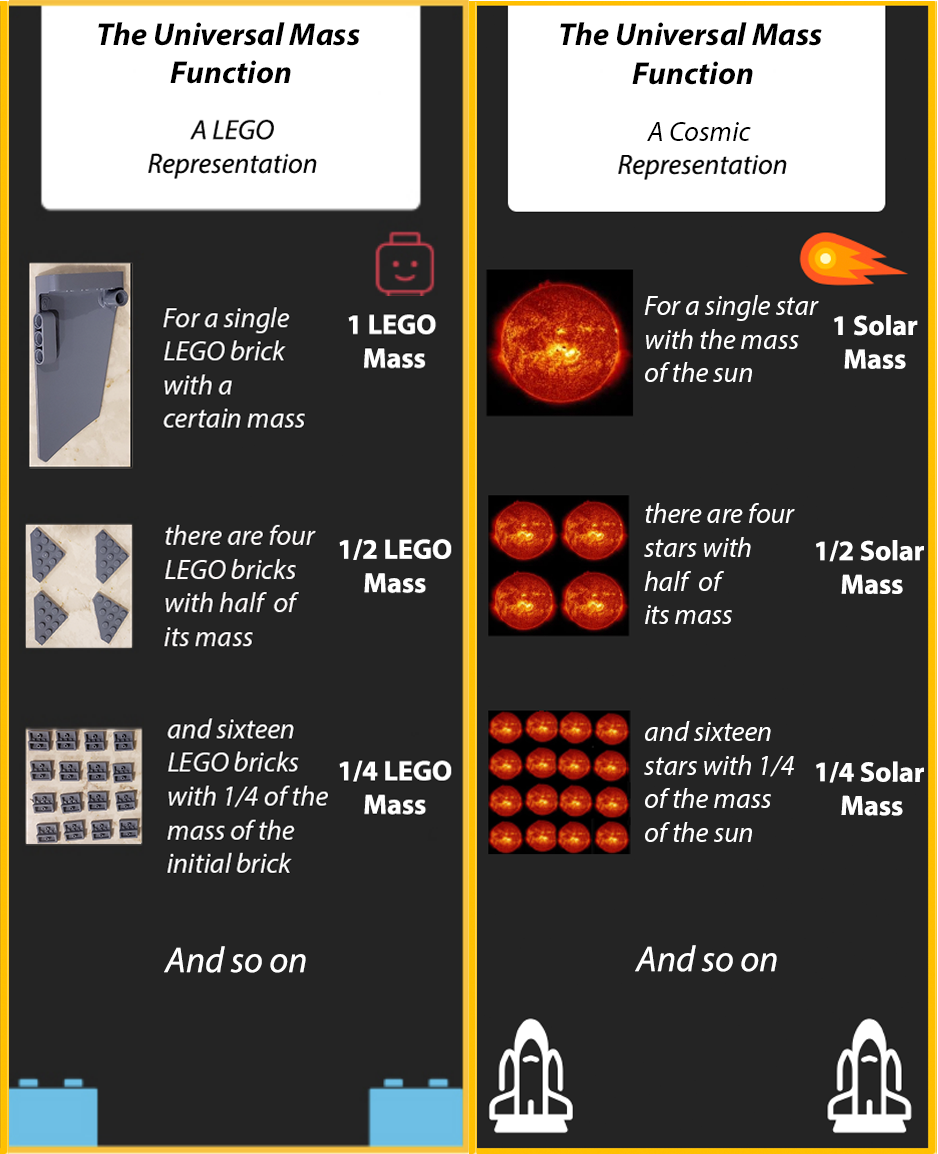}
\caption{This graphic visualizes the meaning of the slope in a mass function and the constant proportionality of mass distributions. We represent the distribution of stars and LEGO bricks from our set using the universal mass function slope of $\alpha = -2$. \label{fig3}}
\end{center}
\end{figure}

\subsection{Other Power Law-like Distributions}

Comprehensive reviews \cite{asc16} \cite[and references therein]{asc18} show that power law-like distributions---similar to mass functions---can be found in many components of nature, such as structure distribution of stellar flares, black holes, planetary surface geometry, galactic structures, in particular impact crater sizes \cite{zan16}, lake sizes \cite{see13}, earthquake magnitudes \cite{pas15}, and many more. This frequency distribution behavior can be also observed in chemistry (e.g., the magnitude distribution of (organic) molecules (Alexenberg \& Layman, private communication), in biology (e.g., the mass distributions of mammals and other animals on Earth: lots of mice, few elephants and whales), in social contexts (e.g., the distribution of wealth, Gini, and Pareto distributions), or just in mass distributions of pieces of broken porcelain, glass, or rocks.  Aschwanden \etal \cite{asc16} \cite{asc18} conclude that these distribution functions follow the concept of self-organizing critical processes (cf. \cite{fri03} \cite{wat16}). Applied to the mass function it could mean the ability of complex systems to self organize (based on the laws of gravitation) on many different mass scales in the Universe.

\subsection{The Origin of Mass Functions in Astronomy}

What is the driving force behind masses to generate the observed mass functions? The origins of individual mass functions are highly debated. For instance, IMFs alone depend on complex interactions between dynamical, chemical, and other environmental properties of star-forming regions \cite{cha03} \cite{off16}. 
Furthermore, the origin of the universal mass function seems to be even more mysterious, because it includes such a wide range of objects in space. Additionally, the general formation processes of different object types (planetary objects, stars, gas and dust clouds, galaxies, galaxy clusters, etc.) are often opposite in nature. For example, stars form by top-down fragmentation of molecular clouds during their gravitationally-driven collapse, while galaxies grow bottom-up via hierarchical gravitational clustering \cite{bin07}.

Mass is the common denominator here. Masses---no matter if considering regular matter, anti matter, or dark matter---interact with each other by the force of gravitation. Therefore, Binggeli \& Hascher \cite{bin07} and Kroupa \etal \cite{kro13} conclude that gravitation is responsible for the continuous universal mass function.  

One could imagine this in the following gedankenexperiment: The gravitational force ($F=G~m_{1}m_{2}/r^{2}$, $G$...gravitational constant) between two point masses ($m_{1}$ and $m_{2}$ separated by the displacement $r$) decreases by the factor of $\frac{1}{r^{2}}$. So, during fragmentation, this $r^{-2}$ dependency of the force means that it is more difficult for fragments to break up from a mass when they are close to a gravitational center, e.g., inside a molecular cloud. But weaker gravitational attraction (and more volume at the outer parts of the cloud) helps to create many more low-mass objects when they are more distant and less attracted from that center. Similarly, clustering affects masses more strongly, when they are close to a gravitational seed and start clumping more effectively compared to loosely attracted, distant masses. Only when mass scales become small, then dynamical, chemical (and electromagnetic) processes start to compete more effectively with gravitation. This is why we see a changing mass function slope typically for low-mass objects.

Consequently, gravitation could explain both, the distribution of mass and the common slope in the universal mass function. The slope of the mass function determines the observed frequency distributions shaped by gravitation. This might be reflected in a connection of the indices in $r^{\alpha}$ of the universal law of gravitation and $m^{\alpha}$ of the universal mass function with the common value of $\alpha = -2$.

\section{Conclusions}

Our experiment provides a pedagogical tool for presenting the concept of mass functions. We also discuss the potential origin of mass functions in space. LEGO bricks are ideal to qualitatively visualize this concept. Moreover, our experiment also reflects the quantitative distributions of masses in the Universe. Therefore, this experiment can inspire students to learn more how the cosmos works and how structures are assembled in the Universe.

\ack

This project was generously funded by the President's Faculty Research and Development Grant 335510 of Nova Southeastern University. DV gratefully acknowledges the support of the STFC via an Ernest Rutherford Fellowship (grant ST/P003850/1). We would like to thank Prof. Dr. D. Castano (Nova Southeastern University, U.S.A.) for assisting us in the work process through intellectual conversations. The plots were made with gnuplot, \verb"http://www.gnuplot.info". The Sun image is from NASA. The comet, head, brick, and shuttle visuals are from \verb"https://Piktochart.com". LEGO is a trademark of The LEGO Group,  \verb"https://www.lego.com/en-us/aboutus/lego-group/".  


{}

\end{document}